\documentclass[twocolumn,showpacs,prb,superscriptaddress]{revtex4}

\usepackage{graphicx}
\usepackage{color}
\usepackage{tabularx}
\usepackage{epsfig}
\usepackage{amsmath}
\usepackage{amssymb}
\usepackage{graphicx}
\usepackage{dcolumn}
\usepackage{bm}
\usepackage{wasysym}

\definecolor{grn}{rgb}{0,0,0.54}

\newcommand{\expect}[1]{\langle \rangle}

\begin{document}

%%%%%%%%%%%%%%%%%%%%%%%%%%%%%%%%%%%%%%%%%%%%%%%%% 

\title{Quantum Phase Diagram and Excitations for the One-Dimensional $S=1$ Heisenberg Antiferromagnet
with Single-Ion Anisotropy}

\author{A. Fabricio Albuquerque}
\affiliation{School of Physics, The University of New South Wales, Sydney, NSW 2052, Australia}

\author{Chris J.~Hamer}
\affiliation{School of Physics, The University of New South Wales, Sydney, NSW 2052, Australia}

\author{Jaan Oitmaa}
\affiliation{School of Physics, The University of New South Wales, Sydney, NSW 2052, Australia}

%%%%%%%%%%%%%%%%%%%%%%%%%%%%%%%%%%%%%%%%%%%%%%%%% 

\date{\today}

\begin{abstract}
We investigate the zero-temperature phase diagram of the one-dimensional $S=1$ Heisenberg
antiferromagnet with single-ion anisotropy. By employing high-order series expansions and quantum
Monte Carlo simulations we obtain accurate estimates for the critical points separating different
phases in the quantum phase diagram. Additionally, excitation spectra and gaps are obtained.
\end{abstract}

\maketitle

%%%%%%%%%%%%%%%%%%%%%%%%%%%%%%%%%%%%%%%%%%%%%%%%%

%####################################################################################

\section{Introduction}
\label{sec:intro}

Interest in one-dimensional $S=1$ antiferromagnets is long-standing and can be traced back to
the original work by Haldane.\cite{haldane:83a,haldane:83b} By analyzing the presence of
topological terms in effective field-theories for one-dimensional antiferromagnets he conjectured
that integer-spin chains display a ground-state with exponentially decaying spin-spin
correlations and a gapped excitation spectrum, properties markedly different from those
displayed by the exactly solvable $S=1/2$ chain. Despite early controversy, this so called
``Haldane conjecture" is now supported by solid numerical\cite{white:93} and
experimental\cite{regnault:94,kenzelmann:02} evidence (for a review of earlier results,
see Ref.~\onlinecite{affleck:89}).

We are interested in the anisotropic $S=1$ antiferromagnetic chain described by the Hamiltonian
\begin{equation}
    {\mathcal H} = 	J \sum_{i} \vec{S}_{i} \cdot \vec{S}_{i+1} - D \sum_{i} ({S}^{z}_{i})^2~.
  \label{eq:hamiltonian}
\end{equation}
The single-ion anisotropy term proportional to $D$ is relevant in accounting for the magnetic
properties of a number of compounds: ${\rm CsNiCl_3}$ (weak axial anisotropy,
Refs.~\onlinecite{steiner:87} and \onlinecite{kenzelmann:02}), NENP
[${\rm Ni(C_{2}H_{8}N_{2})_{2}NO_{2}(ClO_{4})}$] (weak axial anisotropy, Ref.~\onlinecite{renard:88}),
${\rm CsFeBr_3}$, NENC [${\rm Ni(C_{2}H_{8}N_{2})_{2}Ni(CN)_{4}}$] and DTN
[${\rm NiCl_{2}-4SC(NH_{2})_{2}}$] (strong planar anisotropy, Refs.~\onlinecite{dorner:88},
\onlinecite{orendac:95} and \onlinecite{zapf:06} respectively); DTN is particularly interesting due to
Bose-Einstein condensation of spin excitations under a magnetic field.\cite{zapf:06,giamarchi:08}
The model defined by Eq.~(\ref{eq:hamiltonian}) is also very appealing from a purely theoretical perspective:
the single-ion anisotropy $D/J$ controls the magnitude of quantum fluctuations and stabilizes different
phases that, along with the quantum critical points separating them, have been extensively investigated.
\cite{botet:83,glaus:84,schulz:86,papanicolaou:90,sakai:90,golinelli:92,chen:93,chen:03,boschi:03,
boschi:04,venuti:06,tzeng:08a,tzeng:08b}

For large {\em positive} values of the single-ion anisotropy $D/J$ the system is in an Ising-like
N\'{e}el phase characterized by finite staggered magnetization at zero temperature $T=0$
and displaying a gap for holon-like excitations, that we will discuss later in Sec.~\ref{sec:HN}.
Comparatively less well understood is the so-called large-$D$ phase for large {\em negative} values
of $D/J$, first investigated in detail by Papanicolaou and Spathis.\cite{papanicolaou:90} The ground-state
in the large-$D$ phase smoothly evolves from the ground-state for $D/J \rightarrow -\infty$, simply given
by the tensor-product of $S_{i}^{z}=0$ states at each site. The lowest-energy excitations in this phase reside
in the $S_{T}^{z}=\pm 1$ sector and were termed {\em excitons} and {\em anti-excitons} in
Ref.~\onlinecite{papanicolaou:90}, where the existence of bound-states was also verified.

Physical intuition was gained about the intermediate phase observed for small values of $|D|/J$,
including the isotropic point $D/J=0$ analyzed by Haldane,\cite{haldane:83a,haldane:83b} by
studying an extended $S=1$ model with biquadratic interactions
\begin{equation}
    {\mathcal H}_{\rm AKLT} = J \sum_{i} \left[ \vec{S}_{i} \cdot \vec{S}_{i+1} +
    \beta (\vec{S}_{i} \cdot \vec{S}_{i+1})^2 \right]~.
  \label{eq:aklt}
\end{equation}
Affleck, Kennedy, Lieb and Tasaki (AKLT)\cite{affleck:87} showed that this Hamiltonian
is exactly solvable at $\beta = 1/3$, where it displays a simple valence-bond solid (VBS)
ground-state with gapped excitations. den Nijs and Rommelse\cite{dennijs:89}
later suggested that this VBS state can be interpreted as a ``fluid" with positional disorder and long-range
antiferromagnetic order, characterized by a finite expectation value for the string-operator,
\begin{equation}
    {\mathcal O}_{s}(r) = {S}^{z}_{0} \exp \left( i \pi \sum_{k=1}^{r - 1} {S}^{z}_{k} \right) {S}^{z}_{r} ~,
  \label{eq:string}
\end{equation}
in the limit $r \rightarrow \infty$, associated with the breaking of a hidden $\mathbb{Z}_2 \times \mathbb{Z}_2$
symmetry.\cite{kennedy:92}  Since the ground-state at $\beta=0$ has been shown\cite{white:93} to
exhibit long-ranged string correlations and is adiabatically connected to the ground-state at $\beta=1/3$
(see e.g.~Ref.~\onlinecite{schollwoeck:96}), one concludes that the Haldane phase has a VBS character.
We remark that interest in Haldane-like phases exhibiting long-range string correlations has been
renewed and proposals for investigating string-order in cold atomic systems have recently been
made.\cite{garcia-ripoll:04,dallatorre:06} 

In this paper, we are primarily interested in improving on previous estimates for the location of
the large-$D$---Haldane (DH) and the Haldane---N\'{e}el (HN) critical points. Improved results
may be used as benchmarks for further tests on the applicability of quantum information tools
to detect quantum phase transitions in the model Eq.~(\ref{eq:hamiltonian}).\cite{tzeng:08a,tzeng:08b}
Additionally, we obtain results for the excitation spectra and gaps in the large-$D$ and
N\'{e}el phases that may be of experimental relevance. We also show that a series expansion
method previously applied for calculating the Haldane gap\cite{singh:96} yields an incorrect result,
and we propose a different approach for future work.

%####################################################################################

%####################################################################################
\section{Methods}
\label{sec:methods}

We have investigated the $S=1$ antiferromagnetic chain described by the Hamiltonian
Eq.~(\ref{eq:hamiltonian}) by combining high-order series expansions and quantum Monte Carlo
(QMC) simulations. Technical details are described in this Section.

%%%%%%%%%%%%%%%%%%%%%%%%%%%%%%%%%%%%%%%%%%%%%%%%%
\subsection{Series Expansions}
\label{sec:Series}

Numerical linked-cluster expansions have been extensively employed in the investigation
of quantum magnets (for detailed accounts the reader is referred to Refs.~\onlinecite{gelfand:00}
and \onlinecite{oitmaa:06}; a closely related method is discussed in Refs.~\onlinecite{knetter:00a}
and \onlinecite{knetter:03a}) and have been recently derived by some of us\cite{oitmaa:08} for the
square lattice version of the Hamiltonian Eq.~(\ref{eq:hamiltonian}). Among the method's many
appealing features we highlight its applicability to the study of excitations and dynamical responses,
a notoriously difficult task within alternative numerical approaches, following the procedure originally
devised by Gelfand.\cite{gelfand:96}

Basically, the linked-cluster method relies on a suitable decomposition of the lattice Hamiltonian
under investigation,
\begin{equation}
    {\mathcal H} = 	{\mathcal H}_{\rm 0} + \lambda {\mathcal V}~.
  \label{eq:h0V}
\end{equation}
It is assumed that the ground-state of the unperturbed Hamiltonian ${\mathcal H}_{\rm 0}$ is
known and can be written as a tensor product of local states. One proceeds by deriving a
standard Rayleigh-Schr\"{o}dinger perturbative expansion for {\em connected} clusters comprised
of an increasingly large number of sites and, at each step, one subtracts contributions
from embedded subclusters containing fewer sites. In this way, long series for ground-state quantities and
excited states can be exactly calculated up to a certain order in the expansion parameter $\lambda$,
that are then analyzed by means of any suitable extrapolation technique (in this paper we adopt
a standard Pad\'{e} analysis). In what follows, we present the different expansions employed in the
present study.

%%%%%%%%%%%%%%%%%%%%%%%%%%%%%%%%%%%%%%%%%%%%%%%%%
\subsubsection{Large-$D$ Expansion}
\label{sec:LDE}

\begin{figure}
  \begin{center}
    \includegraphics*[width=0.33\textwidth,angle=270]{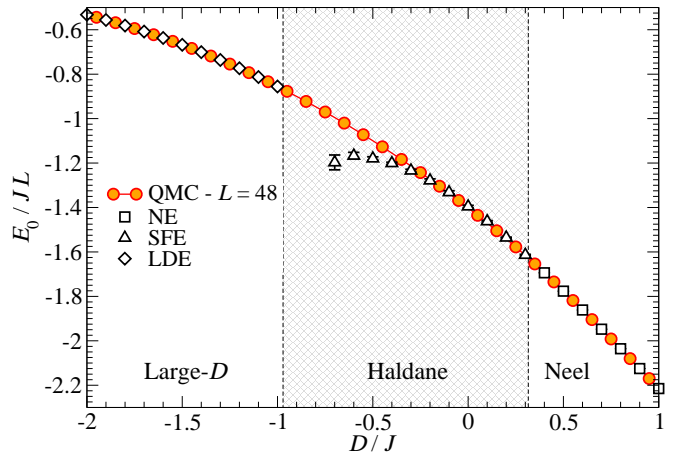}
   \end{center}
  \caption{(Color online) Ground-state energy density for the model described by Eq.~(\ref{eq:hamiltonian})
  as a function of the single-ion anisotropy $D/J$. QMC results are shown as filled circles and
  have been obtained for a lattice with $L=48$ sites.
  Empty symbols correspond to results from the series expansions: large-$D$ expansion (LDE, diamonds),
  N\'{e}el expansion (NE, squares) and staggered-field expansion (SFE, triangles). The vertical dashed lines
  highlight our best estimates for the location of the critical points: $D_{\rm C}^{\rm DH} / J = -0.971(5)$ for the point
  separating the large-$D$ and Haldane phases, and $D_{\rm C}^{\rm HN} / J = 0.316(2)$ for the HN
  phase transition (see main text). Unless visible, error bars are smaller than depicted symbols.}
  \label{fig:E0}
\end{figure}

For investigating the phase stabilized for large {\em negative} values of $D/J$ we
consider the single-ion term as our unperturbed Hamiltonian
\begin{equation}
    {\mathcal H}^{\rm LDE}_{\rm 0} = - D \sum_{i} ({S}^{z}_{i})^2~,
  \label{eq:h0_LDE}
\end{equation}
the superexchange terms in Eq.~(\ref{eq:hamiltonian}) being treated as the perturbation
${\mathcal V}^{\rm LDE}$
\begin{equation}
    {\mathcal V}^{\rm LDE} = J \sum_{i} \left[ S^{z}_{i} S^{z}_{i+1} + \frac{1}{2} \left( S^{+}_{i} S^{-}_{i+1} + 
    						S^{-}_{i} S^{+}_{i+1}  \right) \right]~.
  \label{eq:V_LDE}
\end{equation}
The (non-degenerate) ground-state of ${\mathcal H}^{\rm LDE}_{\rm 0}$
is simply given by tensor products of $S_{i}^{z} = 0$ states for all sites $i$ on the chain and,
in what follows, we refer to this expansion as the {\em large-$D$ expansion} (LDE). Results
for the ground-state energy density obtained from an LDE series comprising terms of up to
$\lambda^{30}$ are shown in Fig.~\ref{fig:E0} (open diamonds). Shorter series have been
obtained for the excited states (up to $\lambda^{14}$), the results being presented in
Sec.~\ref{sec:HD}.

%%%%%%%%%%%%%%%%%%%%%%%%%%%%%%%%%%%%%%%%%%%%%%%%% 
%%%%%%%%%%%%%%%%%%%%%%%%%%%%%%%%%%%%%%%%%%%%%%%%%

%%%%%%%%%%%%%%%%%%%%%%%%%%%%%%%%%%%%%%%%%%%%%%%%%
\subsubsection{N\'{e}el Expansion}
\label{sec:NE}

A suitable linked-cluster expansion for large {\em positive} values of $D/J$ is obtained
by choosing the unperturbed Hamiltonian
\begin{equation}
    {\mathcal H}^{\rm NE}_{\rm 0} = J \sum_{i} S^{z}_{i} S^{z}_{i+1} - D \sum_{i} ({S}^{z}_{i})^2~,
  \label{eq:h0_NE}
\end{equation}
and starting from a perfectly ordered N\'{e}el unperturbed ground-state. The remaining fluctuation terms
in Eq.~(\ref{eq:hamiltonian}) are our perturbation ${\mathcal V}^{\rm NE}$
\begin{equation}
    {\mathcal V}^{\rm NE} = \frac{J}{2} \sum_{i} \left( S^{+}_{i} S^{-}_{i+1} + S^{-}_{i} S^{+}_{i+1} \right)~,
  \label{eq:V_NE}
\end{equation}
and we refer to the resulting expansion as the {\em N\'{e}el expansion} (NE). The ground-state energy and
the staggered magnetization in the N\'{e}el phase have been calculated from a NE comprising terms of to
$\lambda^{18}$ (results for the ground-state energy are depicted as open squares in Fig.~\ref{fig:E0}).
Series for the single-particle excitations have been obtained from expansions up to $\lambda^{14}$;
the results are discussed in Sec.~\ref{sec:HN}.

%%%%%%%%%%%%%%%%%%%%%%%%%%%%%%%%%%%%%%%%%%%%%%%%%
\subsubsection{Staggered-Field Expansion}
\label{sec:SFE}

We have also considered a series expansion originally devised by Singh,\cite{singh:96}
here dubbed as the {\em staggered-field expansion} (SFE), obtained by considering a
Hamiltonian including an artificial staggered-field term
\begin{equation}
\begin{split}
    {\mathcal H}^{\rm SFE} = J \sum_{i} \left[ S^{z}_{i} S^{z}_{i+1}+  \frac{\lambda}{2} \left( S^{+}_{i} S^{-}_{i+1}
    + S^{-}_{i} S^{+}_{i+1} \right) \right] \\
    - D \sum_{i} ({S}^{z}_{i})^2 + (1-\lambda)\sum_{i}(-1)^{i} {S}^{z}_{i} ~.
   \end{split}
  \label{eq:V_SFE}
\end{equation}
(Terms proportional to $\lambda$ are our perturbation ${\mathcal V}^{\rm SFE}$ and the remaining
ones define the unperturbed Hamiltonian ${\mathcal H}_{0}^{\rm SFE}$.)
The unperturbed ground-state is a perfectly ordered N\'{e}el state, as in the case of
the NE, and the Hamiltonian Eq.~(\ref{eq:hamiltonian}) is recovered in the limit $\lambda=1$,
where the last term in Eq.~(\ref{eq:V_SFE}) vanishes. This staggered-field term is included
in order to prevent the condensation of solitons before $\lambda=1$, as discussed in Ref.~\onlinecite{singh:96},
allowing one to obtain a seemingly precise estimate for the Haldane gap at $D/J=0$. Results
for the ground-state energy per site obtained from an SFE including terms up to $\lambda^{18}$
are represented by open triangles in Fig.~\ref{fig:E0}: it is clear from this plot that the SFE systematically
underestimates $E_{\rm 0}/L$ towards the DH critical point and we will address this
issue later in Sec.~\ref{sec:Gap}, along with our results for the gap at $D/J=0$.

\subsection{Quantum Monte Carlo}
\label{sec:QMC}

\begin{figure}
  \begin{center}
    \includegraphics*[width=0.33\textwidth,angle=270]{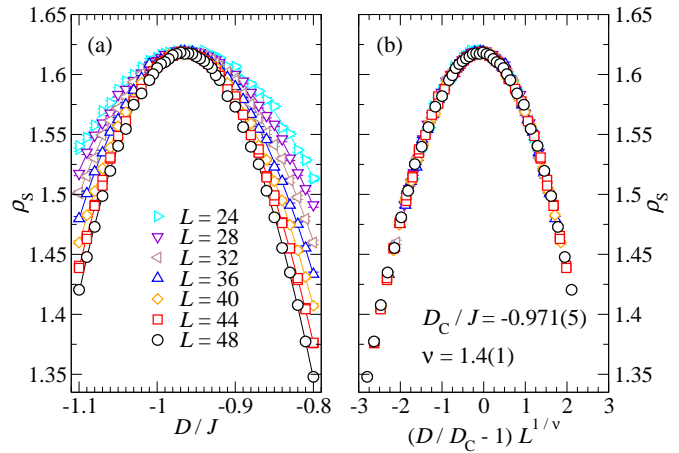}
   \end{center}
  \caption{(Color online) (a) Spin-stiffness $\rho_{\rm S}$ as a function
  of the single-ion anisotropy close to the critical point separating the large-$D$ and Haldane
  phases (lines are only guides to the eye). Results have been obtained from QMC simulations
  for values of $L$ ranging from $24$ to $48$ and error bars are smaller than depicted symbols.
  (b) Data collapse is achieved for $D_{\rm C}^{\rm DH} / J = -0.971(5)$ and $\nu = 1.4(1)$
  (see main text).}
  \label{fig:QMC_DH}
\end{figure}

QMC simulations have been performed by employing the ALPS
(Algorithms and Libraries for Physics Simulations) libraries' \cite{albuquerque:07}
implementation of the {\em directed loops} algorithm \cite{syljuasen:02,alet:05:a} in
the Stochastic Series Expansion (SSE) framework.\cite{sandvik:99} We have
simulated the model Eq.~(\ref{eq:hamiltonian}) on lattices with length $L$ ranging
from 24 to 48, applying periodic boundary conditions. In order to assess ground-state
properties, simulations were performed for temperatures $T = 1/100 < 1/2L$: results for
the spin-stiffness $\rho_{\rm S}$ (Fig.~\ref{fig:QMC_DH}) and the Binder cummulant
$Q_2$ [Eq.~(\ref{eq:binder}), Fig.~\ref{fig:QMC_HN}] are converged, within
statistical errors, to their ground-state expectation values for this value of $T$, as
verified by running preliminary QMC simulations for various temperatures.
Results for the ground-state energy density $E_{\rm 0} / L$ for a lattice $L=48$ are shown
in Fig.~\ref{fig:E0}: the data for $E_{\rm 0} / L$ are essentially converged for this
system size (within statistical errors, as we can conclude by comparing against
results obtained from smaller lattice sizes, not shown here), and good agreement
is obtained with the results from the LDE and NE described in the previous
subsection.

%%%%%%%%%%%%%%%%%%%%%%%%%%%%%%%%%%%%%%%%%%%%%%%%% 
%%%%%%%%%%%%%%%%%%%%%%%%%%%%%%%%%%%%%%%%%%%%%%%%%

%####################################################################################

\section{Numerical Results}
\label{sec:results}

We present estimates for the location of the quantum critical points in the
phase diagram of the Hamiltonian Eq.~(\ref{eq:hamiltonian}) and results for the
excitation spectra and gaps obtained by using the numerical procedures described in
the previous section.

\subsection{Large-$D$ --- Haldane Phase Transition}
\label{sec:HD}

The Gaussian quantum phase transition between the large-$D$ and the Haldane phases
has been studied in a number of recent works.\cite{chen:03,boschi:03,tzeng:08a,
tzeng:08b} The critical point was pinpointed by Chen {\em et al.}\cite{chen:03}~
by performing exact diagonalizations of small clusters with twisted boundary conditions.
Furthermore, these authors verified that the DH transition is described
by a conformal field-theory with central charge $c=1$, as further confirmed by Boschi {\em et al.}
\cite{boschi:03} using a combined field-theoretic and numerical analysis.
More recently, Tzeng and collaborators,\cite{tzeng:08a,tzeng:08b} motivated by the current
interest in applying concepts and tools from quantum information theory to the study of
condensed matter systems,\cite{amico:08} showed that the DH phase transition can be located
by investigating the scaling behavior of the ground-state fidelity and the entanglement entropy,
and arrived at the estimate\cite{errors} $D_{\rm C}^{\rm DH} / J = 0.97$. Here, we investigate the DH transition
employing the numerical methods described in Sec.~\ref{sec:methods}.

\subsubsection{Spin Stiffness}
\label{sec:stiff}

In Fig.~\ref{fig:QMC_DH}(a) we show our QMC results for the spin-stiffness $\rho_{\rm S}$, obtained
in terms of the winding number $w$, $\rho_{\rm S}=3\langle w^2 \rangle / 2 \beta$ ($\beta$ is the inverse
temperature),\cite{sandvik:97} for system sizes $L$ ranging from 24 to 48. Away from the critical point,
both the large-$D$ and the Haldane phases display exponentially decaying spin correlations
and therefore we expect $\rho_{\rm S}$ to approach zero in the thermodynamic limit, a trend
clearly discernible in Fig.~\ref{fig:QMC_DH}(a). On the other hand, close to the critical point, where
the correlation length $\xi$ diverges, the spin-stiffness is expected\cite{sachdev:99a} to scale as $\rho_{\rm S}
\sim L^{2-z-d}$. Since $d=1$ is the spatial dimension and the dynamic critical
exponent is expected to be $z=1$ (see below), one expects $\rho_{\rm S}$ to assume a size independent
value at the transition point (similarly to what happens for the $S=1/2$ chain with easy-plane
anisotropy, see Ref.~\onlinecite{laflorencie:01}), as confirmed by our QMC data plotted in
Fig.~\ref{fig:QMC_DH}(a). We remark that this ``peaked behavior" for $\rho_{\rm S}$ is a signature
of a transition between two phases with exponentially decaying spin correlations and it should be
contrasted with the ``crossing behavior" for $\rho_{\rm S}L$ observed in more conventional order-disorder
quantum transitions in two-dimensional systems (see e.g.~Refs.~\onlinecite{wang:06,wenzel:08,
albuquerque:08c}).

%%%%%%%%%%%%%%%%%%%%%%%%%%%%%%%%%%%%%%%%%%%%%%%%% 
\begin{figure}
  \begin{center}
    \includegraphics*[width=0.32\textwidth,angle=270]{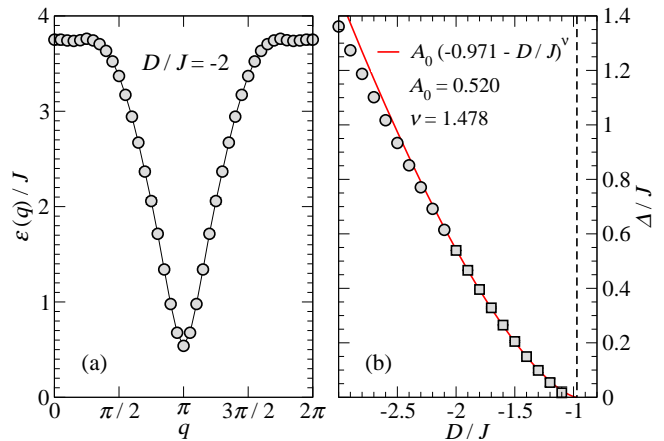}
  \end{center}
  \caption{(Color online) (a) Excitation spectrum in the large-$D$ phase, for $D / J = -2$, obtained from the
  LDE. (b) Gap for excitations ($q=\pi$) in the large-$D$ phase as a function of $D/J$, again obtained from
  the LDE (the series become badly behaved beyond the rightmost data point). The continuous line has been
  obtained by fitting the scaling function $\Delta / J = A_{\rm 0} [(D_{\rm C}^{\rm DH} - D)/J)]^{z\nu}$, assuming
  $z=1$, to the data points depicted as squares, using the QMC estimate $D_{\rm C}^{\rm DH} / J = -0.971$
  (vertical dashed line). In both panels, with the exception of the rightmost data point in (b), error bars are smaller
  than the depicted symbols.}
  \label{fig:LargeD}
\end{figure}
%%%%%%%%%%%%%%%%%%%%%%%%%%%%%%%%%%%%%%%%%%%%%%%%%

In order to estimate the location of the quantum critical point $D_{\rm C}^{\rm DH} / J$ and the correlation
length critical exponent $\nu$, we assume the scaling ansatz
\begin{equation}
  \rho_{\rm S}(t,L) = f_{\rho_{\rm S}} (tL^{1/\nu})~,
  \label{eq:scaling}
\end{equation}
with reduced coupling $t = (D/D_{\rm C}^{\rm DH} - 1)$. Data collapse is achieved for
$D_{\rm C}^{\rm DH} / J = -0.971(5)$ and $\nu=1.4(1)$ [Fig.~\ref{fig:QMC_DH}(b)],
values in good agreement with the ones from Ref.~\onlinecite{tzeng:08a}
($D_{\rm C}^{\rm DH} / J = -0.97$ and $\nu = 1.42$, $1.45$).\cite{errors}
We use our result $\nu=1.4(1)$ to calculate the Luttinger-liquid parameter $K$ by employing the
relation\cite{boschi:03,tzeng:08a,tzeng:08b} $\nu = 1/(2-K)$ and obtain $K=1.29(5)$,
consistent with previous findings.\cite{boschi:03,tzeng:08a,tzeng:08b} Additionally,
the bulk spin-stiffness at the critical point is estimated to be $\rho_{\rm S}^{\rm DH} = 1.619(5)$.

\subsubsection{Excitations}
\label{sec:excitons}

The lowest-lying excitations in the large-$D$ phase lie in the $S^{z}_{\rm T}=\pm 1$ sectors and we
calculate their dispersion relation by applying the LDE presented in Sec.~\ref{sec:LDE}. Results
obtained after a standard Pad\'{e} analysis are shown in Fig.~\ref{fig:LargeD}(a) for $D/J=-2$. As
we approach the DH transition
from the large negative $D/J$ side,
the gap at $q = \pi$ drops to zero, and the dispersion in that neighbourhood becomes linear,
a trend already noticeable for $D/J=-2$ [Fig.~\ref{fig:LargeD}(a)] and consistent with a dynamic
critical exponent $z=1$. We remark that this behavior is not reproduced in the strong-coupling analysis
by Papanicolaou and Spathis,\cite{papanicolaou:90} which gives a quadratic dispersion relation (this is also
the case in our results for couplings deep into the large-$D$ phase, not shown here).
The dependence of the gap at $q=\pi$ on $D/J$ is shown in Fig.~\ref{fig:LargeD}(b). Unfortunately,
the series convergence becomes irregular beyond the rightmost data point in the figure, preventing us from obtaining
independent estimates for $D_{\rm C}^{\rm DH} / J$ and for the critical exponents. In order to partially circumvent
this problem, we fit the scaling function $\Delta / J = A_{\rm 0} [(D_{\rm C}^{\rm DH} - D)/J)]^{z\nu}$, assuming $z=1$,
to the data points depicted as squares in Fig.~\ref{fig:LargeD}(b). By fixing $D_{\rm C}^{\rm DH} / J = -0.971$, as
estimated from our QMC data, the data is fitted for $\nu=1.478$, a value consistent with the QMC result
$\nu=1.4(1)$ and that further confirms that indeed $z=1$.

\subsection{Haldane --- N\'{e}el Phase Transition}
\label{sec:HN}

The N\'{e}el phase for large {\em positive} values of $D/J$ has a twofold degenerate ground-state
and therefore the HN quantum phase transition is expected to belong to the universality class of the
two-dimensional Ising model. Chen {\em et al.}\cite{chen:03} located the HN critical point by
applying a phenomenological renormalization group analysis to data from exact diagonalization
of clusters with up to 16 sites. More recently the estimate\cite{errors} $D_{\rm C}^{\rm HN}/J = 0.31$
was obtained by Tzeng and Yang,\cite{tzeng:08a} who studied the scaling behavior of the ground-state
fidelity close to the HN critical regime.

%%%%%%%%%%%%%%%%%%%%%%%%%%%%%%%%%%%%%%%%%%%%%%%%% 
\begin{figure}
  \begin{center}
    \includegraphics*[width=0.33\textwidth,angle=270]{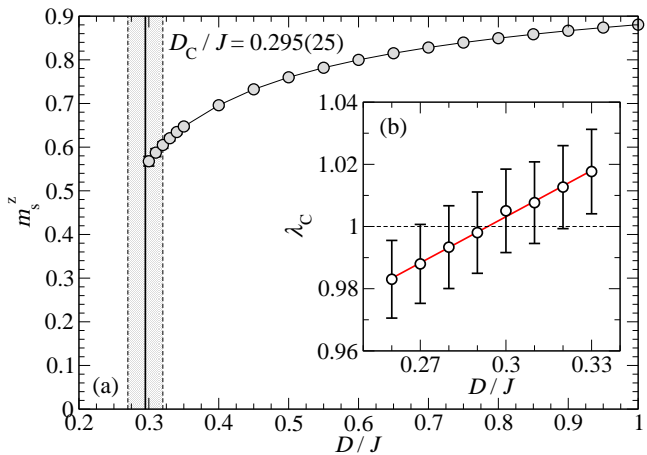}
  \end{center}
  \caption{(Color online) (a) Staggered magnetization per site $m^{z}_{\rm s}$ as a function of $D/J$ in the
  N\'{e}el phase, obtained from the NE by applying a Pad\'{e} analysis. (b) Location of the poles
  $\lambda_{\rm C}$ in the Dlog Pad\'{e} approximants for the staggered magnetization series, as a
  function of $D/J$. 
  In both panels, lines are only guides to the eye.}
  \label{fig:Staggered}
\end{figure}
%%%%%%%%%%%%%%%%%%%%%%%%%%%%%%%%%%%%%%%%%%%%%%%%%
%%%%%%%%%%%%%%%%%%%%%%%%%%%%%%%%%%%%%%%%%%%%%%%%% 
\begin{figure}
  \begin{center}
    \includegraphics*[width=0.4\textwidth]{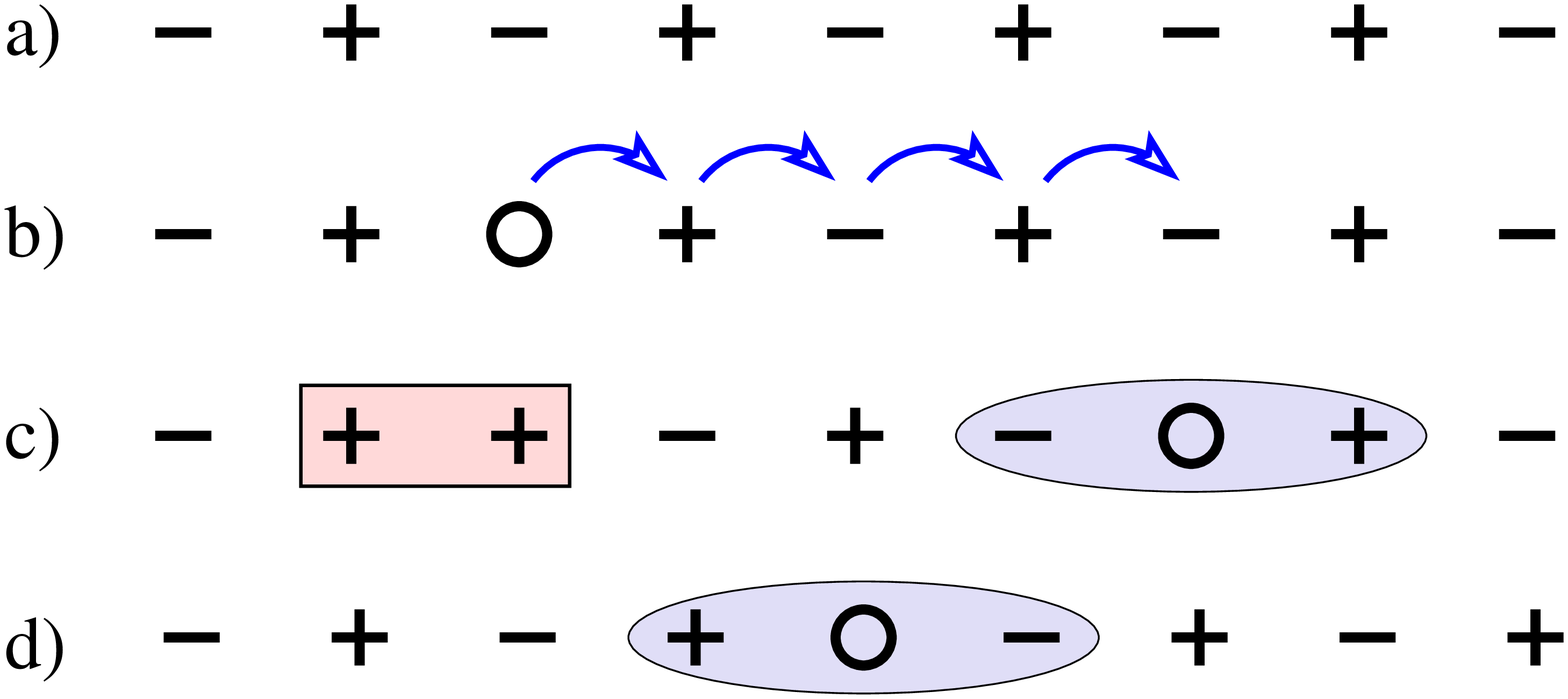}
   \end{center}
  \caption{(Color online) Pseudo-particle representation of the $S=1$ chain.\cite{dennijs:89}
  Sites with $S^{z}_{i}=\pm 1$ are seen as being occupied
  by spin-half particles with pseudo-spin components $\tilde{S}^{z}_{i}=\pm 1/2$ (depicted as $+$'s and
  $-$'s) and sites with $S^{z}_{i}=0$ as being empty (occupied by ``holes", represented by circles)
   In this language, the N\'{e}el ground-state corresponds to an undoped antiferromagnet (a). Excited
   states are obtained by doping the system with holes (b). As depicted in (c), a hole doped into the system decays
   into ``spinon" (enclosed by the rectangle) and holon (enclosed by the ellipsis) constituents [the state depicted
   in (c) is obtained from the state in (b) by allowing the hole to ``hop" four sites to the left, as indicated by the arrows].
   Since a spinon costs an energy $\sim J$, excited states containing only a holon (d) are expected to have lower
   energy (see main text).
   }
  \label{fig:spin_charge}
\end{figure}
%%%%%%%%%%%%%%%%%%%%%%%%%%%%%%%%%%%%%%%%%%%%%%%%%

\subsubsection{Staggered Magnetization}
\label{sec:staggered}

We first employ the NE discussed in Sec.~\ref{sec:NE} in order to calculate the staggered
magnetization, $m^{z}_{\rm s}=\langle (-1)^{i} S^{z}_{i} \rangle / L$, as a function of $D/J$.
The results obtained by applying standard Pad{\' e} approximant extrapolations to the series in
$\lambda$ are shown in Fig.~\ref{fig:Staggered}(a). The position of the HN quantum critical point
can be determined by  a Dlog Pad{\' e} analysis of $m^{z}_{\rm s}$ as a function of $\lambda$:
as shown in Fig.~\ref{fig:Staggered}(b), the estimate $D_{\rm C}^{\rm HN}/J = 0.295(25)$ [highlighted
as the shaded region in Fig.~\ref{fig:Staggered}(a)] for the HN critical point is simply obtained as the
range of values for $D/J$ consistent with a pole at $\lambda=1$, where the full Hamiltonian
Eq.~(\ref{eq:hamiltonian}) is recovered. From the Pad{\' e} extrapolation we also obtain an
estimate\cite{exponents} $\beta = 0.147(13)$ for the critical exponent associated to $m^{z}_{\rm s}$,
which is somewhat larger than, but not incompatible with, the exact result $\beta = 1/8$ for the 2D Ising
universality class. We also note that there is little sign of $m^{z}_{\rm s}$ vanishing in the shaded region in
Fig.~\ref{fig:Staggered}(a): this can be explained by the small value of the expected
critical exponent $\beta = 1/8$, which implies that $m^{z}_{\rm s}$ plunges steeply to zero at the
critical point, a behavior which naive Pad{\' e} approximants will hardly pick up.

\subsubsection{Excitations}
\label{sec:neel_exc}

More accurate estimates for $D_{\rm C}^{\rm HN}/J$ can be obtained by analyzing excited states
above the N\'{e}el ground-state. At this point, following den Nijs and Rommelse,\cite{dennijs:89}
it is useful to interpret the $S=1$ chain as a diluted system of $\tilde{S}=1/2$ pseudo-particles: as we show
in Fig.~\ref{fig:spin_charge}, sites with $S^{z}_{i}=\pm 1$ are seen as being occupied by spin-half
particles with pseudo-spin components $\tilde{S}^{z}_{i}=\pm 1/2$ and sites where $S^{z}_{i}=0$ as being
empty (occupied by ``holes"). Using this language, the N\'{e}el ground-state is equivalent to an ``undoped"
antiferromagnet [Fig.~\ref{fig:spin_charge}(a)] and, for small (positive) values of $D/J$, the low-lying
excited states lie in the ``one-hole sector" (containing {\em one} site with $S^{z}_{i}=0$). Interestingly enough,
the situation is reminiscent of spin-charge separation in one-dimensional fermionic systems (see e.g.
Ref.~\onlinecite{dagotto:94}) where a hole doped into the system [as depicted in Fig.~\ref{fig:spin_charge}(b)]
fractionalizes into ``holon" and ``spinon" constituents [respectively enclosed by an ellipsis and a rectangle in
Fig.~\ref{fig:spin_charge}(c)].

Since one spinon (two consecutive sites with the same pseudo-spin component) has an energetic
cost $\sim J$, we expect $S^{z}_{T}=0$ states solely displaying a holon to have a lower excitation energy,
as confirmed by our results below. Such a holon state is depicted in Fig.~\ref{fig:spin_charge}(d): we replace
one pseudo-spin in the N\'{e}el configuration shown in Fig.~\ref{fig:spin_charge}(a) by a hole and flip
{\em all} pseudo-spins to its right. Antiferromagnetic correlations ``across the hole" allow for the
hole to delocalize, lowering the energy of holon excited states.\cite{dagotto:94} Furthermore, these holon
excitations can be seen as precursors of the ground-state in the Haldane phase: as shown by Nijs and
Rommelse,\cite{dennijs:89} VBS states can be interpreted as a {\em fluid} with positional disorder and
long-range antiferromagnetic order. This picture is consistent\cite{dennijs:89} with the holon state depicted
in Fig.~\ref{fig:spin_charge}(d), but not with the holon-spinon excitation shown in Fig.~\ref{fig:spin_charge}(b-c).\cite{vbs}
Therefore, the HN critical point can be determined by locating the value $D_{\rm C}^{\rm HN}/J$ where the gap
for the holon excitations vanishes.

%%%%%%%%%%%%%%%%%%%%%%%%%%%%%%%%%%%%%%%%%%%%%%%%% 
\begin{figure}
  \begin{center}
    \includegraphics*[width=0.3\textwidth,angle=270]{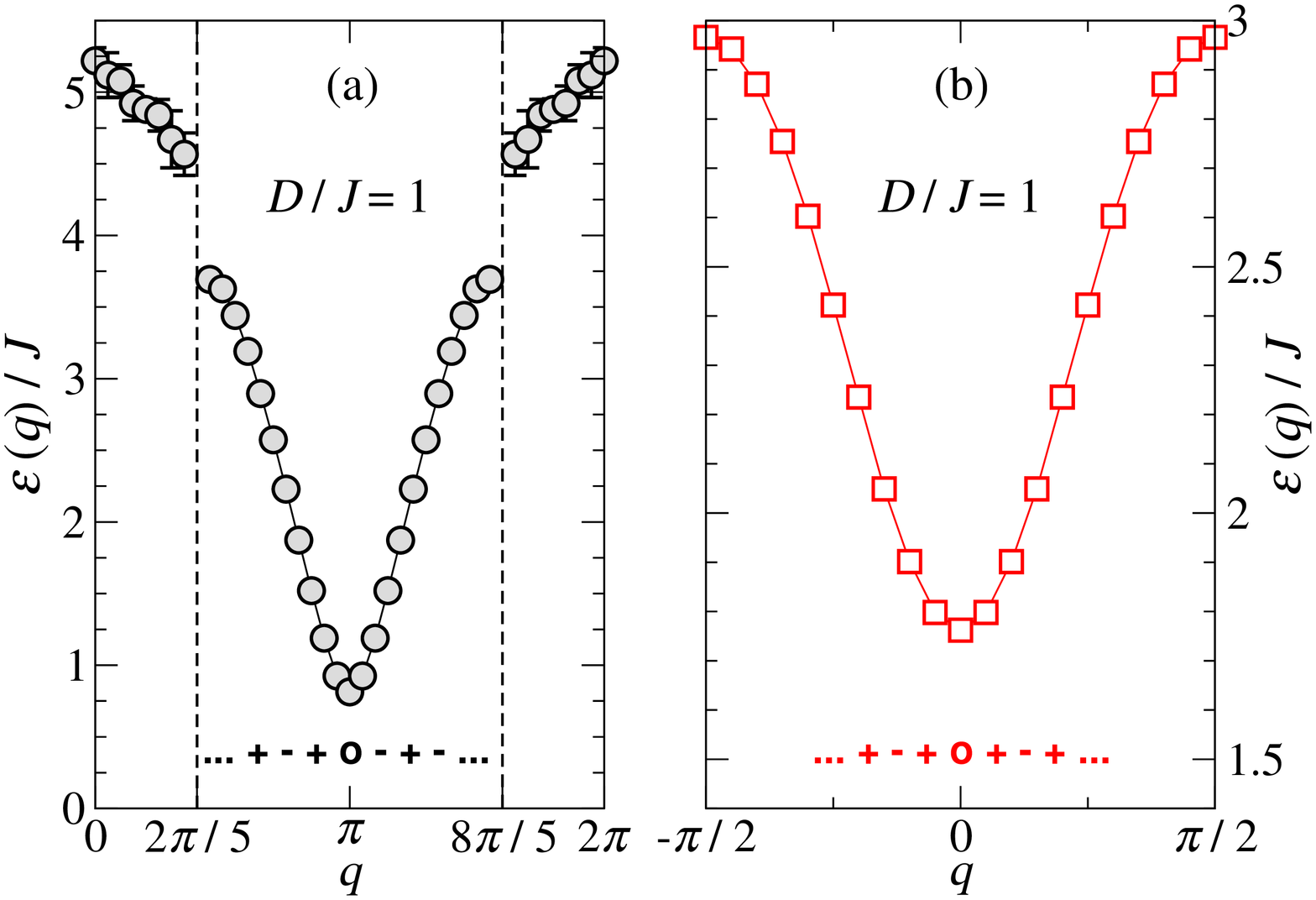}
    
     \vspace{0.3cm}
     
    \includegraphics*[width=0.3\textwidth,angle=270]{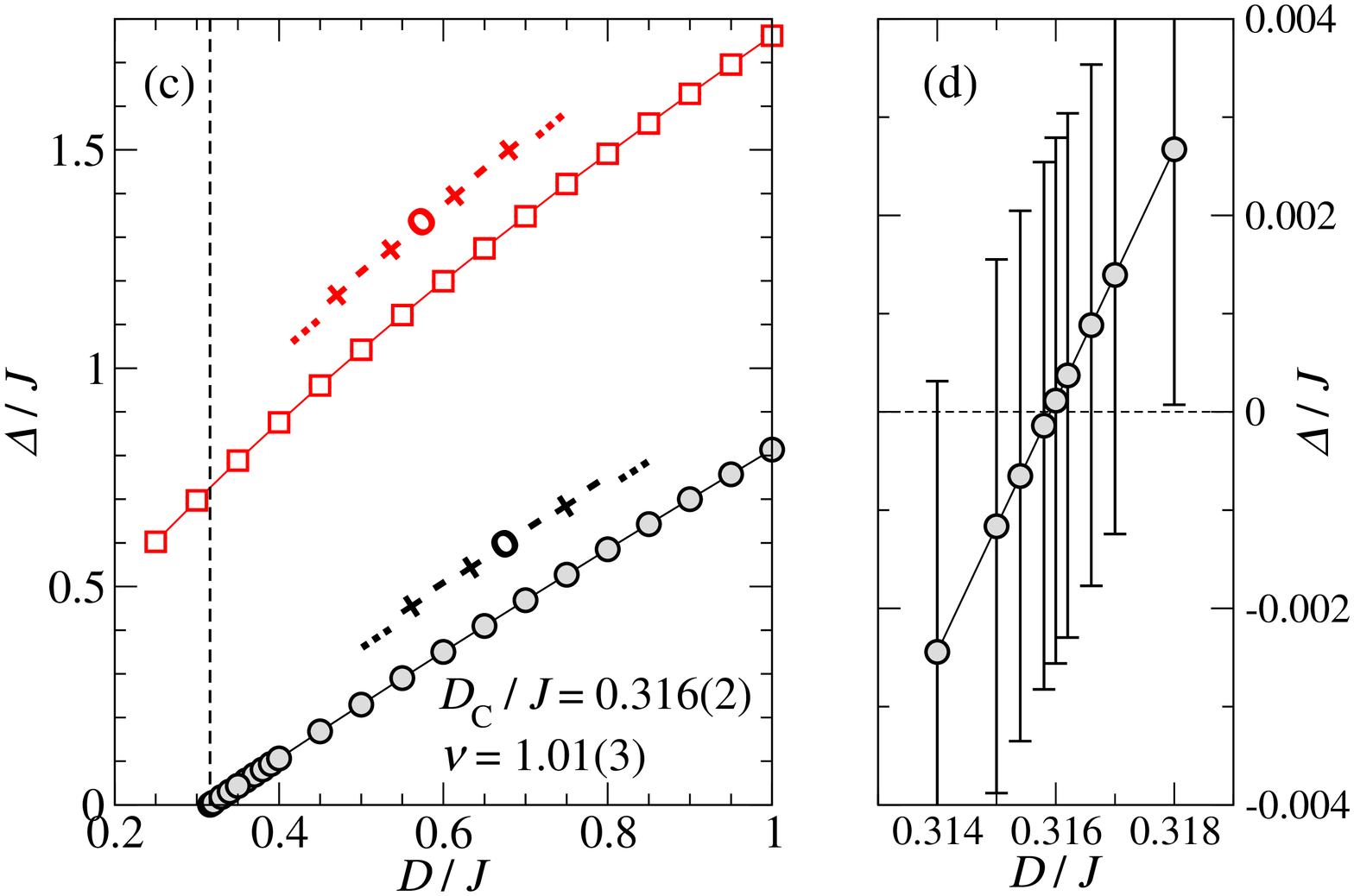}
  \end{center}
  \caption{(Color online) (a) Spectrum for ``holon" excitations [Fig.~\ref{fig:spin_charge}(d)] at $D/J=1$.
  The vertical dashed lines indicate momenta for which the NE expansion seemingly diverges.
  (b) Spectrum for ``holon-spinon" excitations [Fig.~\ref{fig:spin_charge}(b-c)], also for $D/J=1$. (c)
  Excitation gaps as a function of $D/J$. The gap
  for holon excitations ($q=\pi$, black circles) vanishes linearly\cite{exponents} [$\nu=1.01(3)$] at
  $D_{\rm C}^{\rm HN}/J=0.316(2)$ as seen in the magnified plot (d). Data points in all panels were
  obtained by applying a Pad\'{e} analysis to the results obtained from the NE.}
  \label{fig:Neel_Disp}
\end{figure}
%%%%%%%%%%%%%%%%%%%%%%%%%%%%%%%%%%%%%%%%%%%%%%%%%

Results for the spectrum of holon excitations obtained from the NE are shown
in Fig.~\ref{fig:Neel_Disp}(a) for $D/J=1$. We remark that the series seemingly
diverges at the commensurate momenta $q=2\pi/5$ and $q=8\pi/5$ [indicated by the vertical dashed
lines in Fig.~\ref{fig:Neel_Disp}(a)] and that the results at higher energies have a rather poor convergence,
as indicated by the relatively large error bars in the figure. This suggests that holon excitations
decay into multi-holon states and that a continuum of excited states exists at high-energies.
Unfortunately, the poor convergence of the (short) series  for two- and three-holon excited states,
obtained by applying the procedure described in Ref.~\onlinecite{zheng:01}, prevents us from further
analyzing this issue, which is left open to future investigations. On the other hand, our results for low-energy
excitations around $q=\pi$ are nicely converged, allowing us to precisely locate the critical point.
In Figs.~\ref{fig:Neel_Disp}(c-d) we show the dependence of the gap $\Delta/J$ for holon
excitations on $D/J$ and we can see it vanishes at $D_{\rm C}^{\rm HN}/J=0.316(2)$, with a critical
exponent\cite{exponents} $\nu=1.01(3)$ (directly obtained from the Pad\'{e} analysis) consistent with
the 2D Ising universality class.

For the sake of comparison, we also show the dispersion relation for the $S^{z}_{T}=\pm1$
holon-spinon [Fig.~\ref{fig:spin_charge}(b-c)] in Fig.~\ref{fig:Neel_Disp}(b), also for
$D/J=1$. The dependence of the gap at $q=0$ on $D/J$ is shown in Fig.~\ref{fig:Neel_Disp}(c).
Note that our results indicate that the energy of this excitation remains finite at the
transition, confirming that a holon-spinon state has a higher excitation energy than an
isolated holon state. As we see in the plot Fig.~\ref{fig:Neel_Disp}(c), for large values of
$D/J$ the difference in energy between holon-spinon and holon excited states is approximately
$J$, confirming our qualitative analysis above.

%%%%%%%%%%%%%%%%%%%%%%%%%%%%%%%%%%%%%%%%%%%%%%%%% 
\begin{figure}
  \begin{center}
    \includegraphics*[width=0.32\textwidth,angle=270]{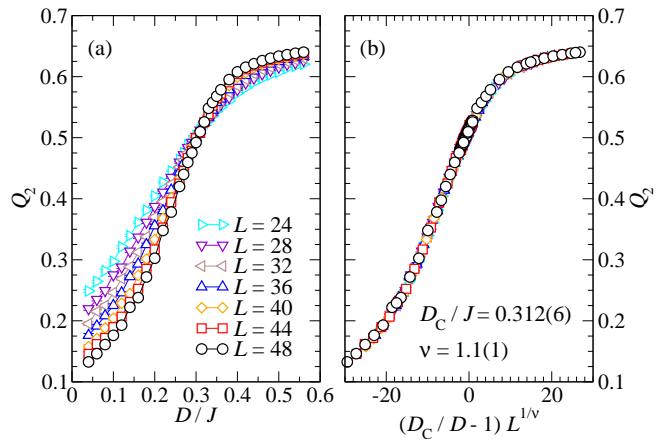}
   \end{center}
  \caption{(Color online) (a) Staggered magnetization Binder cumulant $Q_{\rm 2}$ [see Eq.~(\ref{eq:binder})]
  for couplings $D/J$ close to the Haldane---N\'{e}el critical point, as obtained from QMC simulations for system
  sizes $L$ ranging from 24 to 48. (b) Data collapse is achieved for $D_{\rm C}^{\rm HN}/J=0.312(6)$ and $\nu=1.1(1)$.}
  \label{fig:QMC_HN}
\end{figure}
%%%%%%%%%%%%%%%%%%%%%%%%%%%%%%%%%%%%%%%%%%%%%%%%%

\subsubsection{Binder Cumulant}
\label{sec:binder}

The previous findings are confirmed by our QMC results for the second order
Binder cumulant $Q_2$ for the staggered magnetization $m^{z}_{\rm s}$, given by
\begin{equation}
  Q_{2}= 1 - \frac{\langle (m_{\rm s}^{z})^{4} \rangle}{3 \langle (m_{\rm s}^{z})^{2} \rangle^{2}}~.
  \label{eq:binder}
\end{equation}
$Q_2$ is expected to display universal behavior in the critical regime and results
obtained from different lattice sizes should cross close to the critical point, as
confirmed by the data shown in Fig.~\ref{fig:QMC_HN}(a). In order to estimate the
critical point we assume the scaling ansatz $Q_2(t,L) = f_{Q_2} (tL^{1/\nu})$,
with $t = (D/D_{\rm C}^{\rm HN} - 1)$. Data collapse is achieved for
$D_{\rm C}^{\rm HN}/J = 0.312(6)$ and $\nu=1.1(1)$, as shown in
Fig.~\ref{fig:QMC_HN}(b), consistent with the estimates obtained from the
NE. A slightly smaller value, $D_{\rm C}^{\rm HN}/J = 0.310(6)$, is estimated from
a data collapse obtained by fixing the critical exponent to the known value for the
2D Ising universality class, $\nu=1$.

Additionally, from the data collapse displayed in Fig.~\ref{fig:QMC_HN}(b) we
arrive at the estimate $Q_{2}^{\ast}=0.515(10)$ for the value assumed by
the Binder cummulant at the critical point, markedly different from the
known result\cite{kamieniarz:93} $Q_2^{\ast}=0.61069...$ for the two-dimensional
Ising universality class. This discrepancy can be ascribed to differences
in the way $Q_2$ is calculated in classical and quantum Monte Carlo
simulations: although the $d$-dimensional quantum system
is indeed formally equivalent to a classical model in $d+1$ dimensions,
equal-time expectation values for diagonal operators are averaged along
the imaginary-time direction in quantum simulations. In other words, our
results for $Q_2^{\ast}$ should be consistent with the ones obtained for the
classical two-dimensional Ising model if the moments of the magnetization
appearing in the definition Eq.~(\ref{eq:binder})
are evaluated along individual lines on the square lattice. Since, to the best
of our knowledge, no such calculation has been done, this prevents us
from comparing our estimate $Q_{2}^{\ast}=0.515(10)$ with published results.
We remark that a similar situation was observed in Ref.~\onlinecite{wang:06}
for $S=1/2$ bilayers, expected to belong to the universality class of the
classical Heisenberg model in three dimensions.

\subsection{Haldane Gap}
\label{sec:Gap}

Finally, we have investigated the SFE introduced by Singh\cite{singh:96} and discussed in
Sec.~\ref{sec:SFE} in order to estimate the Haldane gap at $D/J=0$. Singh assumes that the
magnon gap in the N{\' e}el phase of the SFE system will extrapolate continuously to the
Haldane gap as $\lambda \rightarrow 1$, in spite of the phase transition that must occur at or before
$\lambda =1$. The extrapolated value for the magnon gap, $\Delta/J=0.42(1)$,\cite{gap} obtained from
a series involving terms of up to $\lambda^{10}$ as presented in Ref.~\onlinecite{singh:96}, is indeed consistent
with the high-precision result for the Haldane gap $\Delta/J=0.41050(2)$ from density-matrix renormalization
group (DMRG) calculations.\cite{white:93} However, higher-order Pad\'{e} approximants
for an extended series comprising terms up to $\lambda^{14}$ show a clear trend towards
smaller values and we arrive at the considerably lower value $\Delta/J=0.23(6)$.
It is thus obvious that a more careful analysis is required. If we apply a change of variable
$\delta = 1-\sqrt{1-\lambda}$ to the series, following Huse,\cite{huse:88} then Pad{\' e}
approximants for the series in $\delta$ show a vanishing magnon gap,  $\Delta/J=0.00(1)$, at the
transition point $\lambda =1$. The failure to yield correct results for the Haldane gap, combined
with the fact that it systematically underestimates the ground-state energy towards the DH critical point,
as mentioned in Sec.~\ref{sec:SFE} in connection with Fig.~\ref{fig:E0}, strongly suggests that the SFE 
approach is not appropriate for estimating results for the integer-spin Haldane systems.

We recall that a N\'{e}el state is chosen as the unperturbed state when performing the SFE, and the
VBS character of the Haldane phase is ignored. We expect physical results to be obtainable from a
different expansion assuming the AKLT model [$\beta = 1/3$ in Eq.~(\ref{eq:aklt})] as the
unperturbed Hamiltonian and starting from the exact VBS ground-state\cite{affleck:87} at
$\beta = 1/3$, treating $\lambda = (\beta - 1/3)$ as the expansion parameter. Such an
expansion would require the application of the linked-cluster formalism to valence-bond states,
something that, to the best of our knowledge, has not yet been tried and that would constitute an
interesting topic for future work.

%####################################################################################

\section{Conclusions}

Summarizing, we have investigated the one-dimensional $S=1$ antiferromagnet with
single-ion anisotropy, described by the Hamiltonian Eq.~(\ref{eq:hamiltonian}), by means
of linked-cluster series expansions and QMC simulations. Our estimates for the zero-temperature
phase transitions in this model are more precise than previous ones and could be used as
benchmarks in future explorations of the applicability of quantum information tools to the
study of quantum critical phenomena.

Our best estimate for the DH critical point [$D_{\rm C}^{\rm DH}/J=0.971(5)$] has been
obtained from a scaling analysis of the spin-stiffness $\rho_{\rm S}$ from QMC
simulations. The spin-stiffness remains finite at the transition in the thermodynamic limit
[$\rho_{\rm S}^{\rm DH} = 1.619(5)$] and vanishes elsewhere, implying a ``peaked
behavior" for finite systems. Our result for $D_{\rm C}^{\rm DH}$ agrees with the
estimate\cite{errors} obtained by Tzeng et al.\cite{tzeng:08a} [$D_{\rm C}^{\rm DH}=0.97$],
the same being true for the estimate for the correlation length critical exponent:
$\nu = 1.4(1)$ (present work) and $\nu = 1.42$, $1.45$ (Ref.~\onlinecite{tzeng:08a}).
We have further obtained results for the dispersion relation and gap for excitations in
the large-$D$ region that may be of direct experimental relevance given, for instance,
the current interest on the large-$D$ compound DTN.\cite{zapf:06,giamarchi:08} Our
series for the excited states in the large-$D$ phase have a considerably more extended
range of applicability compared to the strong coupling results of Papanicolaou and
Spathis,\cite{papanicolaou:90} and are available on request.

Precise results for the HN phase transition have been obtained from a linked-cluster
expansion (NE) for holon-like excitations. By analyzing the gap for these excited states
we have arrived at the estimates $D_{\rm C}^{\rm HN}/J=0.316(2)$ and $\nu = 1.01(3)$.
The former compares well with the result\cite{errors} $D_{\rm C}^{\rm HN}/J=0.31$ from
Ref.~\onlinecite{tzeng:08a} while the latter confirms that the HN transition belongs to
the universality class of the two-dimensional Ising model, with the exact exponent
$\nu =1$.

Finally, we have shown evidence that the SFE expansion of Singh\cite{singh:96} does not
converge to the Haldane gap at $D/J=0$. Instead, we propose an expansion
around the AKLT state [the exact ground-state for the Hamiltonian Eq.~(\ref{eq:aklt}) with
$\beta = 1/3$],\cite{affleck:87} which we hope to explore in future work.

%####################################################################################

\begin{acknowledgments}
We acknowledge fruitful discussions with O.~P.~Sushkov, N.~Laflorencie and M.~Troyer.
This work has been supported by the Australian Research Council.
\end{acknowledgments}

\bibliographystyle{apsrev}

\begin{thebibliography}{58}
\bibitem[{err()}]{errors}
\bibinfo{note}{The authors of Refs.~\onlinecite{tzeng:08a} and
  \onlinecite{tzeng:08b} do not provide error bars for their different
  estimates.}
\bibitem[{exp()}]{exponents}
\bibinfo{note}{We rely on the assumption that the critical exponents are the
  same when approaching the HN critical point by varying $D/J$ or the expansion
  coupling $\lambda$.}
\bibitem[{gap()}]{gap}
\bibinfo{note}{By applying a Pad\'{e} analysis to the same series we arrive at
  the consistent result $\Delta/J=0.40(3)$.}
\end{thebibliography}

\end{document}